\documentclass[9pt, lettersize]{article}
\usepackage{amsmath,amsfonts}
\usepackage{algorithmic}
\usepackage{array}
\usepackage{cite}
\usepackage[caption=false,font=normalsize,labelfont=sf,textfont=sf]{subfig}
\usepackage{textcomp}
\usepackage{stfloats}
\usepackage{url}
\usepackage{verbatim}
\usepackage{graphicx}
\usepackage[labelfont=bf]{caption}

\def\BibTeX{{\rm B\kern-.05em{\sc i\kern-.025em b}\kern-.08em
    T\kern-.1667em\lower.7ex\hbox{E}\kern-.125emX}}
\usepackage{balance}
\usepackage[margin=1in]{geometry}
\begin{document}
\title{Practical Vectorial Mode Solver for Dielectric Waveguides Based on Finite Differences}
\author{Ergun Simsek}

\maketitle
\begin{abstract}
\textbf{This study presents a finite-difference-based numerical solver designed for the electric field formulation of vector wave equations in optically linear, non-magnetic, dielectric waveguides. We construct a generalized eigenvalue problem by incorporating all three components of the electric field into a self-consistent formulation. This ensures accurate enforcement of boundary conditions and reduces numerical artifacts, particularly at permittivity discontinuities. We validate the solver's performance through two representative waveguide structures, demonstrating its accuracy in computing both propagation constants and mode profiles.}
\end{abstract}
\section{Introduction}
Electromagnetic wave propagation in dielectric structures has been one of the most fundamental subjects of our field. Early studies were dated as old as the beginning of the previous century \cite{EMwavesInDie}.  
After the invention of the laser, the interest in dielectric waveguides surged in the 1960s, and dielectric waveguides became the backbone for the confinement and propagation of optical signals in modern photonics and telecommunication. As a result, several numerical \cite{Ahmed1969, DieWG2Dfem1984, Bierwirth1986, Hadley2002, die_wg_Murphy2008, DieWG2DbiLu14, allmethods, okamoto, 1058177, 4407559, 6022738} and approximate methods \cite{Marcatili1969, Mittra1975} for the modal analysis of optical dielectric waveguides have been proposed. The most common techniques used in these studies are the finite-element method \cite{Ahmed1969, DieWG2Dfem1984}, the finite-difference method \cite{Bierwirth1986, Hadley2002, die_wg_Murphy2008}, and the integral-equation method \cite{DieWG2DbiLu14}. For other methods, the readers are kindly referred to  \cite{allmethods} and \cite{okamoto}.
Among these techniques, finite-element and finite-difference methods have emerged as prominent tools, offering flexibility in handling complex geometries and material properties. However, challenges such as spurious solutions and computational inefficiencies persist, particularly in solving the full vector-wave equation.

\begin{figure}
    \centering
    \includegraphics[width=0.75\linewidth]{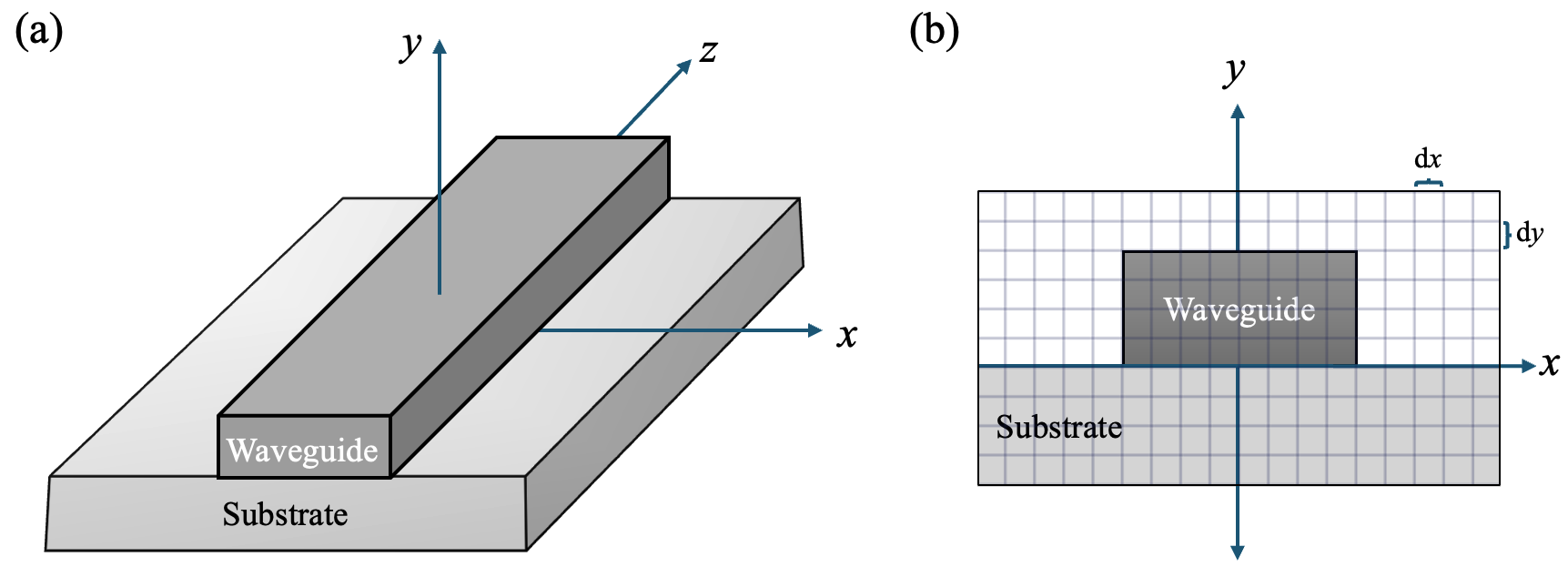}
    \caption{(a) Three-dimensional illustration of a waveguide on top of a substrate. (b) With the uniformity along the $z$-direction, the problem can be simplified into two dimensions and solved on a rectangular grid with mesh sampling densities of d$x$ and d$y$ along the $x$ and $y$ directions. }
    \label{fig:general}
\end{figure}

In this study, we focus on longitudinally uniform dielectric waveguides composed of optically linear, non-magnetic media, aiming to find numerical solutions for the guided modes these waveguides support. Figure \ref{fig:general} illustrates the geometry of our interest. By assuming a time-harmonic field variation $e^{j(\omega t-\beta z)}$, we reduce Maxwell’s source-free equations to vector-wave equations for the electric field. While conventional approaches (e.g., \cite{Bierwirth1986,Hadley2002}) often rely on the full magnetic field formulation to enforce boundary conditions and continuity, the presence of discontinuities in material interfaces and the emergence of spurious solutions remain major hurdles.

To address these issues, we propose a simple yet effective finite-difference-based numerical solver specifically tailored for the electric field formulation. In the following paragraphs, we present the theoretical framework underlying our numerical solver. We then demonstrate the solver’s performance on two representative waveguide problems, showcasing its accuracy in calculating guided modes. 

We should emphasize that the sole purpose of this work is to provide a practical and reliable computational tool for the modal analysis of dielectric waveguides. The enforcement of electromagnetic field boundary conditions and their derivatives has been thoroughly addressed in the literature, particularly in the development of high-order, full-vectorial mode solvers for dielectric waveguides \cite{1058177, 4407559, 6022738}. For instance, Chiou and Du \cite{6022738} demonstrated a highly accurate and computationally efficient eigenmode solver capable of reaching machine precision. In contrast, the goal of our work is to present a simplified and interpretable finite-difference-based implementation suitable for exploratory analysis and integration with data-driven approaches. While our method does not aim for machine-level accuracy, it provides sufficient precision for the targeted class of problems and remains flexible for future extensions. A more advanced treatment of boundary conditions, such as those in \cite{1058177, 4407559, 6022738}, or permittivity averaging \cite{perm_averaging} could be incorporated in future iterations of our framework to improve accuracy where needed.

\section{Formulation}

The formulation starts with the vector wave equation for the electric field
\begin{equation}
\nabla^2\mathbf{E} + \nabla\left(\frac{1}{\varepsilon}\nabla\varepsilon \cdot \mathbf{E}\right) + k_0^2\varepsilon_r\mathbf{E} = 0,
\label{waveeq}
\end{equation}
where $k_0 = \omega\sqrt{\mu_0\varepsilon_0}$ is the free-space wavenumber and $\varepsilon_r = \varepsilon/\varepsilon_0$ is the relative electrical permittivity.

We employ phasor notation and consider a two-dimensional configuration by assuming that the waveguide extends infinitely along the \( z \)-axis. Accordingly, we express the electric field as  
\begin{equation}
\mathbf{E}(x,y,z) = \left[ \hat{x}E_x(x,y) + \hat{y}E_y(x,y) + \hat{z}E_z(x,y) \right] e^{-j\beta z},
\label{Efield}
\end{equation}  
where \( \beta \) denotes the propagation constant in the \( z \)-direction. By substituting \eqref{Efield} into the wave equation \eqref{waveeq}, we derive the coupled equations governing the components \( E_x \), \( E_y \), and \( E_z \).

While the expression for the Laplacian of \( \mathbf{E} \) is standard and can be found in most calculus textbooks, we provide the full form of the term \( \nabla \left( \frac{1}{\varepsilon} \nabla \varepsilon \cdot \mathbf{E} \right) \) below for completeness:
\begin{equation}
\begin{aligned}
\nabla \left( \frac{1}{\varepsilon} \nabla \varepsilon \cdot \mathbf{E} \right) = & \left\{
\hat{\mathbf{x}} \frac{\partial}{\partial x} \left[ \frac{1}{\varepsilon} \left( \frac{\partial \varepsilon}{\partial x} E_x + \frac{\partial \varepsilon}{\partial y} E_y \right) \right] \right. \\
& + \hat{\mathbf{y}} \frac{\partial}{\partial y} \left[ \frac{1}{\varepsilon} \left( \frac{\partial \varepsilon}{\partial x} E_x + \frac{\partial \varepsilon}{\partial y} E_y \right) \right] \\
& \left. + \hat{\mathbf{z}} \left[ -j\beta \frac{1}{\varepsilon} \left( \frac{\partial \varepsilon}{\partial x} E_x + \frac{\partial \varepsilon}{\partial y} E_y \right) \right] \right\} e^{-j\beta z}.
\end{aligned}
\label{Term2}
\end{equation}

The resulting wave equation can be decomposed into three scalar equations corresponding to the \( x \)-, \( y \)-, and \( z \)-components of the electric field:
\begin{equation}
\frac{\partial^2 E_x}{\partial x^2} + \frac{\partial^2 E_x}{\partial y^2} 
+ \frac{\partial}{\partial x} \frac{1}{\varepsilon} \left( \frac{\partial \varepsilon}{\partial x} E_x + \frac{\partial \varepsilon}{\partial y} E_y \right)
+ k_0^2\varepsilon_r E_x = \beta^2 E_x,
\label{Eeqs1}
\end{equation}
\begin{equation}
\frac{\partial^2 E_y}{\partial x^2} + \frac{\partial^2 E_y}{\partial y^2} 
+ \frac{\partial}{\partial y} \frac{1}{\varepsilon} \left( \frac{\partial \varepsilon}{\partial x} E_x + \frac{\partial \varepsilon}{\partial y} E_y \right) 
+ k_0^2\varepsilon_r E_y = \beta^2 E_y,
\label{Eeqs2}
\end{equation}
\begin{equation}
\frac{\partial^2 E_z}{\partial x^2} + \frac{\partial^2 E_z}{\partial y^2}   
-j\beta \frac{1}{\varepsilon} \left( \frac{\partial \varepsilon}{\partial x} E_x + \frac{\partial \varepsilon}{\partial y} E_y \right)
+ k_0^2\varepsilon_r E_z = \beta^2 E_z.
\label{Eeqs3}
\end{equation}

One might carry out the modal analysis by working only on the \eqref{Eeqs1} and  \eqref{Eeqs2} and then determine the $z$-component with
\begin{equation}
E_z = \frac{1}{j\beta \varepsilon_r} \left( \frac{\partial}{\partial x} \varepsilon_r E_x + \frac{\partial}{\partial y} \varepsilon_r E_y \right).
\label{EzEq}
\end{equation} 
However, this formulation does not inherently ensure the correct behavior of $E_z$ near material interfaces when implemented using a straightforward finite-difference scheme. In particular, numerical inaccuracies can arise at boundaries where discontinuities in permittivity exist due to the presence of spatial derivatives of $\varepsilon$ in the expression for $E_z$. To mitigate these discretization-induced artifacts and obtain a more accurate numerical solution—especially near interfaces—we opt to include all three components of the electric field in our formulation. This choice addresses implementation challenges without implying a fundamental improvement to the underlying electromagnetic model.
To achieve this, we cast \eqref{Eeqs1}, \eqref{Eeqs2}, and \eqref{Eeqs3} into the following matrix equation
\begin{equation}
\begin{bmatrix}
M_1 & M_2 & M_3 \\
M_4 & M_5 & M_6 \\
M_7 & M_8 & M_9
\end{bmatrix}
\begin{bmatrix}
E_x \\
E_y \\
E_z 
\end{bmatrix}
= 
\beta^2 
\begin{bmatrix}
E_x \\
E_y \\
E_z 
\end{bmatrix},
\end{equation}
where $M_3 = M_6 =0$ and 
\begin{align}
        M_1 & = M_9 + \frac{\partial}{\partial x} \frac{1}{\varepsilon} \frac{\partial \varepsilon}{\partial x}, \\
        M_2 & = \frac{\partial}{\partial x} \frac{1}{\varepsilon} \frac{\partial \varepsilon}{\partial y},\\
        M_4 & = \frac{\partial}{\partial y} \frac{1}{\varepsilon} \frac{\partial \varepsilon}{\partial x}, \\
        M_5 & = M_9 + \frac{\partial}{\partial y} \frac{1}{\varepsilon} \frac{\partial \varepsilon}{\partial y}, \\
        M_7 & = -j\beta \frac{1}{\varepsilon} \frac{\partial \varepsilon}{\partial x}, \label{M7} \\
        M_8 & = -j\beta \frac{1}{\varepsilon} \frac{\partial \varepsilon}{\partial y}, \label{M8} \\
        M_9 & =  \frac{\partial^2 }{\partial x^2} + \frac{\partial^2}{\partial y^2} + k_0^2\varepsilon_r.         
\end{align}

If we can determine all the three components of the electric field, then we can find the magnetic field components with the following expressions:
\begin{equation}
H_x = \frac{j}{\omega \mu_0} \left(\frac{\partial E_z}{\partial y} + j\beta E_y\right),     
\label{hx}
\end{equation}
\begin{equation}
H_y = \frac{j}{\omega \mu_0} \left( -j\beta E_x - \frac{\partial E_z}{\partial x} \right),    
\label{hy}
\end{equation}
\begin{equation}
H_z = \frac{j}{\omega \mu_0} \left( \frac{\partial E_y}{\partial x} - \frac{\partial E_x}{\partial y} \right).
\label{hz}
\end{equation}
Note that since the above equations do not include $\varepsilon$, they can be computed more accurately than \eqref{EzEq}. 

The problem in this formulation is that we have the unknown, ``$\beta$", on both sides of \eqref{Eeqs3}. To handle this issue, we write the matrix equation in the following format
\begin{equation}
\overline{\overline{{M}}}\mathbf{E} + \beta \overline{\overline{{L}}}\mathbf{E} - \beta^2 \overline{\overline{{I}}} \mathbf{E} = 0,
\end{equation}
where $\overline{\overline{{M}}}$ is the matrix independent of $\beta$, $\overline{\overline{{L}}}$ is the linear term in $\beta$, and the last term is the quadratic term in $\beta$.  This can be rewritten as a linear generalized eigenvalue problem by introducing an auxiliary variable $\mathbf{F} = \beta \mathbf{E}$, leading to $\overline{\overline{{M}}} \mathbf{E} + \overline{\overline{{L}}} \mathbf{F} - \mathbf{F} \beta = 0,$ which gives the augmented system:
\begin{equation}
    \begin{pmatrix}
        \overline{\overline{{M}}} & \overline{\overline{{L}}} \\
        0 & \overline{\overline{{I}}}
    \end{pmatrix}
    \begin{pmatrix}
        \mathbf{E} \\
        \mathbf{F}
    \end{pmatrix}
    = \beta
    \begin{pmatrix}
        0 & \overline{\overline{{I}}} \\
        \overline{\overline{{I}}} & 0
    \end{pmatrix}
    \begin{pmatrix}
        \mathbf{E} \\
        \mathbf{F}
    \end{pmatrix}.
\end{equation}

\begin{figure}[t]
    \centering
    \includegraphics[width=0.9\linewidth]{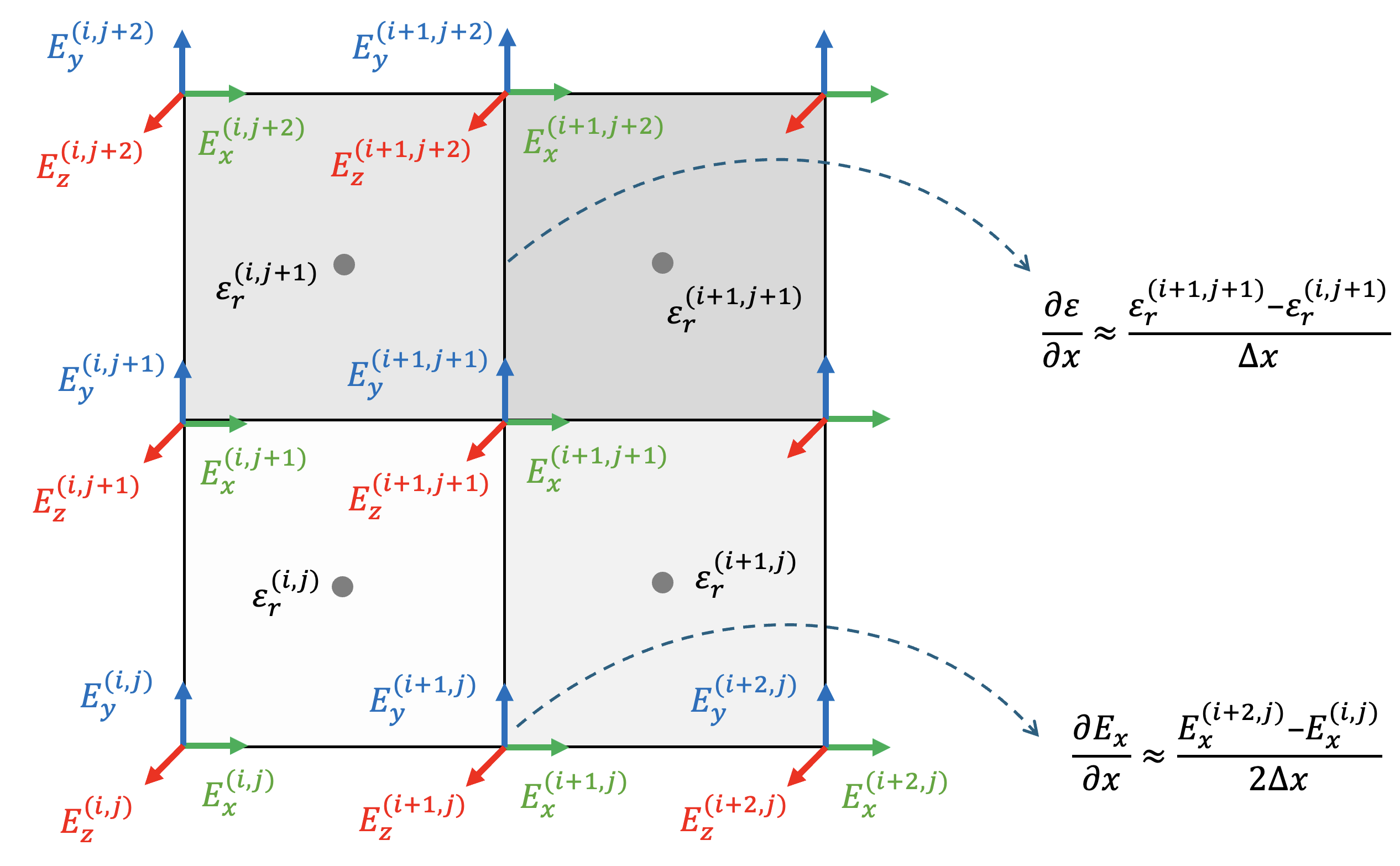}
    \caption{Visualization of the finite-difference grid layout used to approximate spatial derivatives of the electric field components and relative permittivity \(\varepsilon_r\).}
    \label{fig:FDmesh}
\end{figure}

Figure~\ref{fig:FDmesh} illustrates a typical grid used to compute spatial derivatives of the electric field components and the relative permittivity \(\varepsilon_r\) in a finite-difference framework. Each grid cell corresponds to a discrete spatial region, and the electric field components \(E_x\), \(E_y\), and \(E_z\) are evaluated at the cell edges, while the relative permittivity \(\varepsilon_r\) is evaluated at the center of each cell. In the figure, we use first-order accurate finite difference formulas for interoperability. In our numerical implementation, we employ a higher-order scheme with a fourth-order central difference stencil for improved accuracy with the following expressions
\begin{eqnarray}
    \frac{\partial A}{\partial \rho} \approx &  \frac{-A(\rho+2h) + 8A(\rho+h) - 8A(\rho-h) + A(\rho-2h)}{12h}, \\
    \frac{\partial^2 A}{\partial \rho^2} \approx & \frac{-A(\rho+2h) + 16A(\rho+h) - 30A(\rho) + 16A(\rho-h) - A(\rho-2h)}{12h^2},
\end{eqnarray}
where $A$ is either $E_x$, $E_y$, $E_z$, or $\varepsilon$, $h$ is the unit mesh length along the $\rho$ direction, and $\rho$ is either $x$ or $y$.

It is important to note that the spatial derivatives of the permittivity, such as \(\partial \varepsilon_r / \partial x\) or \(\partial \varepsilon_r / \partial y\), are only non-zero at interfaces where material properties change—such as at the boundary between a waveguide and the surrounding cladding. In homogeneous regions, where \(\varepsilon_r\) is constant, these derivatives vanish. On the outermost boundary nodes, both the field values and their spatial derivatives are assumed to be zero, which is a reasonable approximation as long as the computational domain is chosen sufficiently large, since the guided modes are expected to be concentrated within the core of the dielectric waveguide.

We solve the linear generalized eigenvalue problem using MATLAB's \textsf{eigs} function, which is optimized for large sparse matrices. The maximum number of iterations is set to 1000. As an initial guess, we enter $\beta_{\rm{guess}} = k_0 (n_{\rm{ring}} - j10^{-8})$, yielding the first 10 modes in approximately 2–3 seconds with a tolerance of $10^{-12}$. After solving, we retain only the physically meaningful modes (those with $n_{\rm{eff}}<n_{\rm{ring}}$.) The determined eigenvectors are the true distributions of $E_x$, $E_y$, and $E_z$ over the computation domain. We determine the magnetic field components with \eqref{hx}, \eqref{hy}, and \eqref{hz}. As for the final step, we compute the power density and normalize all the fields so that the power density of the electromagnetic wave is equal to 1 W/m$^2$.

\section{Numerical Results}
To demonstrate the accuracy of this method, we provide two numerical examples. In both examples, the excitation wavelength is 1550 nm.

For the first example, we consider a Si$_3$N$_4$ waveguide surrounded by SiO$_2$. Their refractive indices are assumed to be 1.9761 and 1.444. The width and height of the waveguide are 1.6 $\mu$m and 0.7 $\mu$m. The meshing density along the $x$ and $ y$ directions is 20 points per wavelength (PPW). The thickness of the SiO$_2$ coating is 1 $\mu$m in all directions. Table \ref{table_example1} lists the effective refractive indices of the first four resonant modes computed with a commercial solver, Tidy3D, and our solver. Since for the Tidy3D implementation, we use 80 PPW, we take Tidy3D solutions as the ground truth and calculate the difference in percentile as listed in the last column of Table \ref{table_example1}. We observe that the difference is less than 1 \% for all modes.

\begin{table} 
\caption{Effective refractive indices of the first four resonant modes of a $1.6 \ \mu$m$\times0.7 \ \mu$m Si$_3$N$_4$ waveguide surrounded by SiO$_2$, computed with Tidy3D and our numerical solver. The fourth column is the percentage difference between Tidy3D and our solutions. }
\begin{center} 
\begin{tabular}{|c|c|c|c|} 
\hline 
mode & $n_{\mathrm{eff}}$ (Tidy3D) & $n_{\mathrm{eff}}$ (This work) & Difference (\%) \\ 
\hline 
    1  & 1.7951  &  1.7933  &  0.1014  \\
    \hline
    2  &  1.7540  &   1.7507 &   0.1889 \\
    \hline
    3  &  1.6485  &  1.6464  &  0.1290 \\
    \hline
    4  &  1.6294  &  1.6257  &  0.2241\\
    \hline
\end{tabular} 
\end{center} 
\label{table_example1} 
\end{table} 

\begin{figure}[ht]
\centering
\fbox{\includegraphics[width=0.9\linewidth]{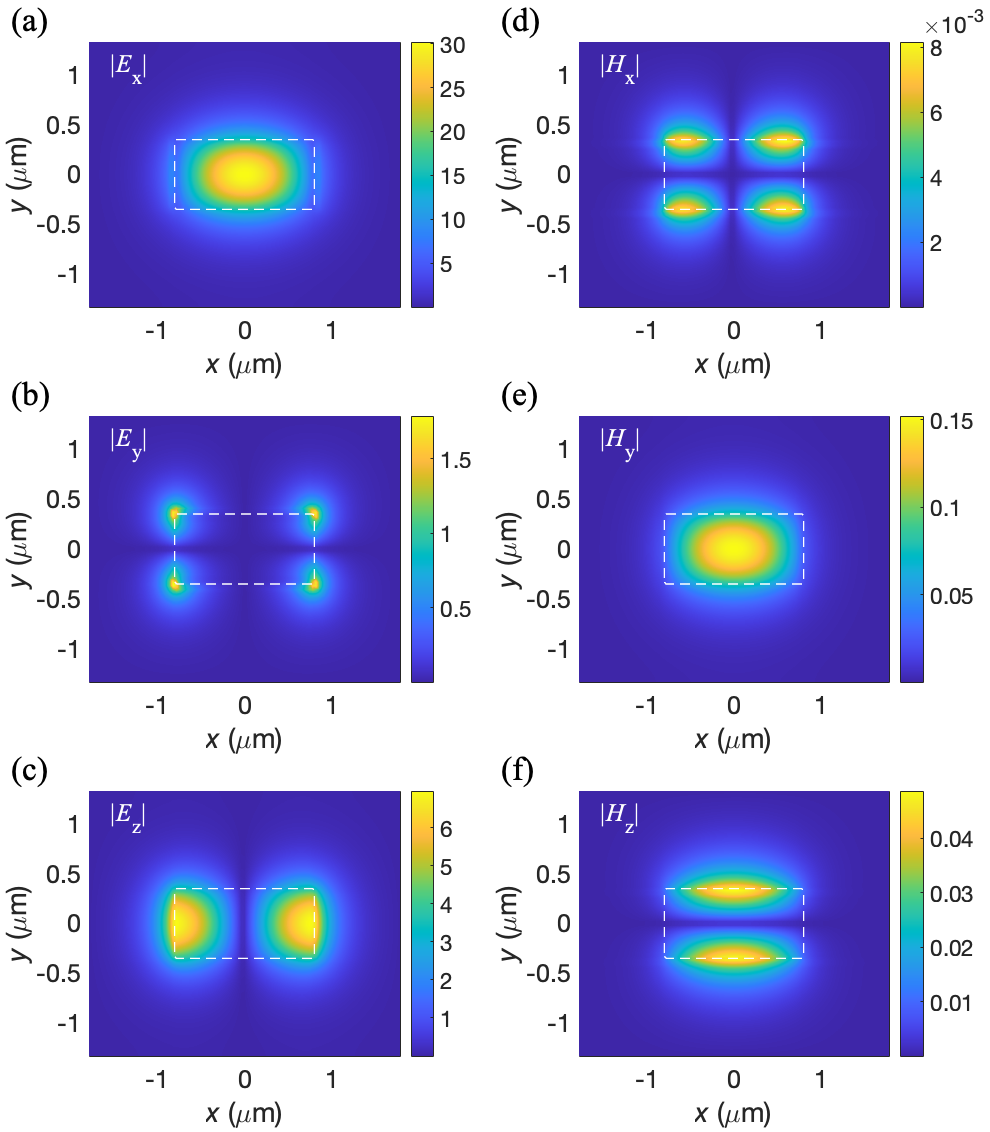}}
\caption{The magnitude of the electric (left) and magnetic (right) field components for the first resonant mode of the electromagnetic wave propagating in a Si$_3$N$_4$ waveguide with a width and height of 1.6 $\mu$m and 0.7 $\mu$m. The white dashed lines outline the boundaries of the waveguide, giving insight into how the electromagnetic fields are confined.}
\label{fig:modes_1}
\end{figure}

Figure \ref{fig:modes_1} shows the magnitude of the $x$, $y$, and $z$ components of the electric and magnetic fields for the first resonant mode. The $|E_x|$ component has a strong intensity concentrated at the center of the waveguide (almost 95 \% of the total electric field), corresponding to the core region. The $|H_y|$ field is also concentrated within the core region, showing a complementary distribution to $|E_x|$. The field distributions determined by the commercial solver are not provided here for brevity, but they are almost identical to those shown in Fig. \ref{fig:modes_1}.

\begin{figure}[ht]
\centering
\fbox{\includegraphics[width=\linewidth]{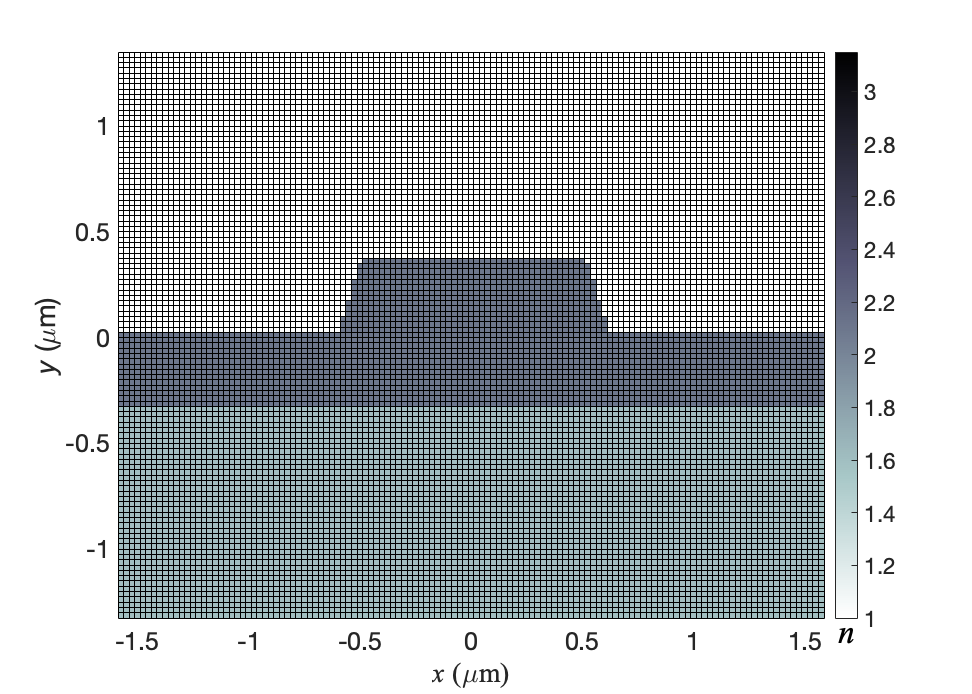}}
\caption{The mesh used in the second example, where the colors represent the refractive index of each mesh element.}
\label{fig:mesh}
\end{figure}

In the second example, we work on a trapezoid waveguide placed on top of a thin film-coated substrate, as illustrated in Fig. \ref{fig:mesh}. The refractive index of the waveguide and thin film is chosen to be 2.1 to mimic thin-film lithium niobate waveguides \cite{TFLN}. The refractive indices of the substrate and the region above the waveguide are 1.6 and 1, respectively. The height of the waveguide and film thickness are 350 nm. The width of the waveguide decreases from 1.2 $\mu$m to 1.0 $\mu$m along the $y$-direction. To capture this varying width, we set the mesh sampling density to 30 PPW, a slightly higher value than the one used in the first example.

Table \ref{table_example2} lists the effective refractive indices of the first four resonant modes computed with Tidy3D and our solver and their differences in percentile. We again observe that the difference is less than 1 \% for all the modes.

\begin{table} 
\caption{Follows the Table \ref{table_example1} for a trapezoid waveguide illustrated in Fig. \ref{fig:mesh}. }
\begin{center} 
\begin{tabular}{|c|c|c|c|} 
\hline 
mode & $n_{\mathrm{eff}}$ (Tidy3D) & $n_{\mathrm{eff}}$ (This work) & Difference (\%) \\ 
\hline 
    1  &  1.8893  &  1.8890 &   0.0170 \\
    \hline
    2  &  1.8344  &  1.8230 &   0.6202 \\
    \hline
    3  &  1.7651  &  1.7497 &   0.8715 \\
    \hline
    4  &  1.6827  &  1.6864 &   0.2179 \\
    \hline
\end{tabular} 
\end{center} 
\label{table_example2} 
\end{table} 

\begin{figure}[ht]
\centering
\fbox{\includegraphics[width=0.9\linewidth]{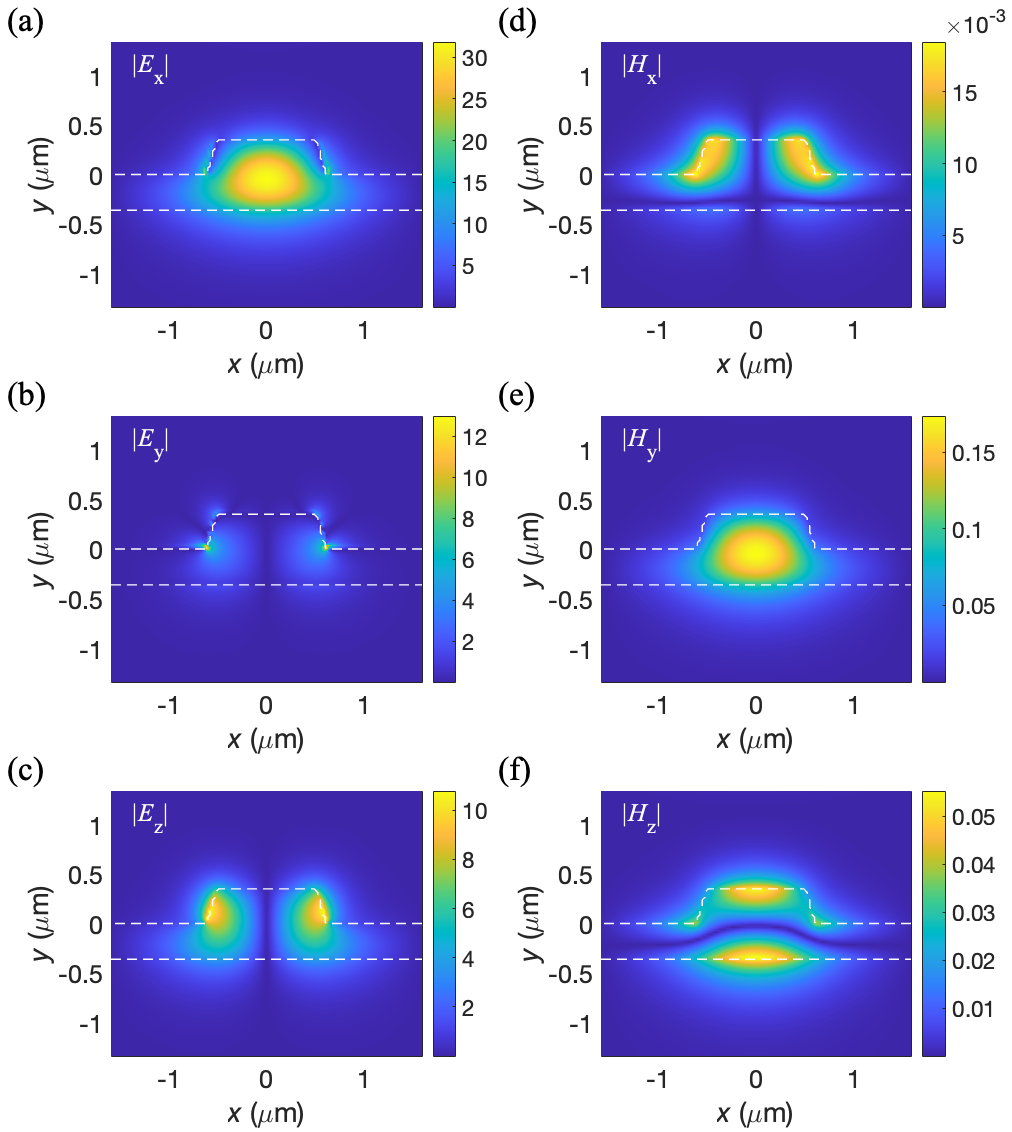}}
\caption{Similar to Fig. \ref{fig:modes_1} for a trapezoid waveguide on a thin film-coated substrate illustrated in Fig. \ref{fig:mesh}.}
\label{fig:modes_2}
\end{figure}

Figure \ref{fig:modes_2} shows the magnitude of the electric and magnetic field components for the first resonant mode. The $|E_x|$ component has a strong intensity concentrated at the center of the waveguide, corresponding to the core region. The profile is symmetric about the $x=0$ plane, consistent with typical guided modes in symmetric structures. The $|E_y|$ component shows intensity peaks near the edges of the waveguide's core, which are associated with field variations across the interface of the waveguide core and the surrounding medium. The $|H_y|$ field is concentrated within the core region, showing a complementary distribution to $|E_x|$.
    
\section{Conclusions}
We have developed a finite-difference-based numerical solver for the electric field formulation of vector wave equations in optically linear, non-magnetic, and dielectric waveguides. By incorporating all three components of the electric field into a unified eigenvalue formulation, our method improves the enforcement of boundary conditions and minimizes numerical artifacts near permittivity discontinuities. This self-consistent approach provides a practical balance between simplicity and accuracy, offering reliable mode solutions without requiring the complexity of high-order schemes. Demonstrations on representative waveguide geometries validate the effectiveness of our solver in capturing both the propagation constants and modal field profiles.
\clearpage
\bibliographystyle{ieeetr}
\bibliography{references}

\begin{thebibliography}{10}

\bibitem{EMwavesInDie}
D.~Hondros and P.~Debye, ``Elektromagnetische wellen an dielektrischen
  dr{\"a}hten,'' {\em Annalen der Physik}, vol.~337, pp.~465--476, 2024/10/13
  1910.

\bibitem{Ahmed1969}
S.~Ahmed and P.~Daly, ``Finite-element methods for inhomogeneous waveguides,''
  {\em Proceedings of the Institution of Electrical Engineers}, vol.~116,
  pp.~1661--1664(3), October 1969.

\bibitem{DieWG2Dfem1984}
B.~Rahman and J.~Davies, ``Finite-element solution of integrated optical
  waveguides,'' {\em Journal of Lightwave Technology}, vol.~2, no.~5,
  pp.~682--688, 1984.

\bibitem{Bierwirth1986}
K.~Bierwirth, N.~Schulz, and F.~Arndt, ``Finite-difference analysis of
  rectangular dielectric waveguide structures,'' {\em IEEE Transactions on
  Microwave Theory and Techniques}, vol.~34, no.~11, pp.~1104--1114, 1986.

\bibitem{Hadley2002}
G.~Hadley, ``High-accuracy finite-difference equations for dielectric waveguide
  analysis i: uniform regions and dielectric interfaces,'' {\em Journal of
  Lightwave Technology}, vol.~20, no.~7, pp.~1210--1218, 2002.

\bibitem{die_wg_Murphy2008}
A.~Fallahkhair, K.~S. Li, and T.~E. Murphy, ``Vector finite difference
  modesolver for anisotropic dielectric waveguides,'' {\em Journal of Lightwave
  Technology}, vol.~26, no.~11, pp.~1423--1431, 2008.

\bibitem{DieWG2DbiLu14}
W.~Lu and Y.~Y. Lu, ``Efficient high order waveguide mode solvers based on
  boundary integral equations,'' {\em Journal of Computational Physics},
  vol.~272, pp.~507--525, 2014.

\bibitem{allmethods}
K.~S. Chiang, ``Review of numerical and approximate methods for the modal
  analysis of general optical dielectric waveguides,'' {\em Optical and Quantum
  Electronics}, vol.~26, no.~3, pp.~S113--S134, 1994.

\bibitem{okamoto}
K.~Okamoto, {\em Fundamentals of optical waveguides}.
\newblock Elsevier, 2010.

\bibitem{1058177}
Y.-C. Chiang, Y.-P. Chiou, and H.-C. Chang, ``Improved full-vectorial
  finite-difference mode solver for optical waveguides with step-index
  profiles,'' {\em Journal of Lightwave Technology}, vol.~20, no.~8,
  pp.~1609--1618, 2002.

\bibitem{4407559}
P.-J. Chiang, C.-L. Wu, C.-H. Teng, C.-S. Yang, and H.-c. Chang,
  ``Full-vectorial optical waveguide mode solvers using multidomain
  pseudospectral frequency-domain (psfd) formulations,'' {\em IEEE Journal of
  Quantum Electronics}, vol.~44, no.~1, pp.~56--66, 2008.

\bibitem{6022738}
Y.-P. Chiou and C.-H. Du, ``Arbitrary-order full-vectorial interface conditions
  and higher order finite-difference analysis of optical waveguides,'' {\em
  Journal of Lightwave Technology}, vol.~29, no.~22, pp.~3445--3452, 2011.

\bibitem{Marcatili1969}
E.~A.~J. Marcatili, ``Dielectric rectangular waveguide and directional coupler
  for integrated optics,'' {\em The Bell System Technical Journal}, vol.~48,
  no.~7, pp.~2071--2102, 1969.

\bibitem{Mittra1975}
W.~McLevige, T.~Itoh, and R.~Mittra, ``New waveguide structures for
  millimeter-wave and optical integrated circuits,'' {\em IEEE Transactions on
  Microwave Theory and Techniques}, vol.~23, no.~10, pp.~788--794, 1975.

\bibitem{perm_averaging}
S.~G. Johnson and J.~D. Joannopoulos, ``Block-iterative frequency-domain
  methods for maxwell's equations in a planewave basis,'' {\em Optics Express},
  vol.~8, no.~3, pp.~173--190, 2001.

\bibitem{TFLN}
B.~Desiatov, A.~Shams-Ansari, M.~Zhang, C.~Wang, and M.~Lon{\v c}ar,
  ``Ultra-low-loss integrated visible photonics using thin-film lithium
  niobate,'' {\em Optica}, vol.~6, no.~3, pp.~380--384, 2019.

\end{thebibliography}
\end{document}